\documentclass[useamsfonts, proof]{pasj00}
\draft

\begin{document}
\SetRunningHead{Jiang et al.}{Microflares and magnetic fields}
%\Received{}%{yyyy/mm/dd}
%\Accepted{}%{yyyy/mm/dd}
%\Published{}%{yyyy/mm/dd}

\title{Relationship between EUV microflares and small-scale magnetic fields in the quiet Sun}

\author{Fayu \textsc{Jiang},  Jun \textsc{Zhang}, Shuhong \textsc{Yang}}
\affil{Key Laboratory of Solar Activity, National Astronomical Observatories, \\Chinese Academy of Sciences, 
Beijing 100012, China}
\email{jiangfayu@nao.cas.cn, zjun@nao.cas.cn, shuhongyang@nao.cas.cn}


\maketitle

\begin{abstract}

Microflares are small dynamic signatures observed in X-ray and extreme-ultraviolet channels.
Because of their impulsive emission enhancements and wide distribution, they are thought to be
closely related to coronal heating. By using the high resolution 171 \AA\ images from the Atmospheric Imaging Assembly and the lines-of-sight magnetograms obtained by the Helioseismic and Magnetic Imager on board the
\emph{Solar Dynamics Observatory}, we trace 10794 microflares in a quiet region near the disk center with a field of view of $960\arcsec \times 1068\arcsec$ during 24 hr. The microflares have an occurrence rate of
$4.4\times10^{3}$~hr$^{-1}$
extrapolated over the whole Sun. Their average brightness, size, and lifetime are 1.7 $I_{0}$ (of the quiet Sun), 9.6 Mm$^{2}$, and 3.6 min, respectively. There exists a mutual positive correlation between the microflares' brightness, area and lifetime. In general, the microflares distribute uniformly across the solar disk, but form network patterns locally, which are similar
to and matched with the magnetic network structures. Typical cases show that the microflares prefer to occur in magnetic cancellation regions of network boundaries.
We roughly calculate the upper limit of energy flux supplied by the microflares and find that the result is still a factor of 
$\sim 15$ below the coronal heating requirement. 
 
\end{abstract}

\section{Introduction}

Bright points (BPs) are small-scale, roundish, dynamic structures observed in X-ray and extreme-ultraviolet (EUV) channels and have caused great concerns since their first discovery in rocket X-ray images by \citet{1973ApJ...185L..47V}. 
\citet{1974ApJ...189L..93G} found the BPs had a spatial size of about 10$^{8}$ km$^{2}$ and a mean lifetime of 8 hr 
from the \textit{Skylab} X-ray images. \citet{1981SoPh...69...77H} observed BPs in EUV channels and found the BPs showed large 
intensity variations in a few minutes. The BPs are widespread throughout the whole solar atmosphere, e.g., in the corona \citep{2013SoPh..286..125C}, the photosphere \citep{2007ApJ...661.1272B} and the transition region \citep{1997SoPh..175..467H}. They are observed in active regions (e.g., \cite{1995PASJ...47..251S, 2001A&A...369..291B, 2012SoPh..280..407R}), coronal holes \citep{1990ApJ...352..333H, 2010ApJ...710.1806D, 2011A&A...529A..21K}, and quiet regions \citep{1990ApJ...352..333H, 2001SoPh..198..347Z, 2011A&A...529A..21K}. Being observed by different instruments in different wavelengths, they
have been given many different names, for example,
active-region transient brightenings, blinkers, coronal BPs, microflares, nanoflares,
X-ray BPs, X-ray jets, etc. (see \cite{2002ESASP.505..231P, 2004psci.book.....A}).
Although their physical nature is not fully understood, these small-scale phenomena may share similar physical processes.
BPs are found to be associated with dipolar magnetic fields \citep{1974ApJ...189L..93G}, and the corresponding total flux
surpass that of active regions \citep{1997soco.book.....G}.

The massive energy involved in the BPs, together with their widespread distribution among the solar surface, make small transient events heating mechanism a promising candidate in solving coronal heating problem.  
The idea of small reconnection events responsible for 
coronal heating was
proposed by \citet{1974ApJ...190..457L}, and then developed by \citet{1988ApJ...330..474P}
who
suggested that the magnetic foot-point motion  could form many small current sheets 
heating the corona via Ohmic dissipation.  From balloon-borne observations, \citet{1984ApJ...283..421L} discovered many
small hard X-ray microflares with energies between $10^{24} - 10^{27}$ ergs. \citet{1987ApJ...323..380P} suggested the 
BPs observed in the EUV line, which lied above the neutral lines of dipoles, 
were microflares and drivers of spicules.  \citet{1991SoPh..133..357H} pointed out that if the power low index of energies distribution $\alpha$ $> -2$, large flares dominated the corona heating, otherwise small flares dominated.  Various studies have 
suggested the power law index lied in $-2.7 < \alpha < -1.5$ (see \cite{2012RSPTA.370.3217P}). 
Statistical study of microflares in active regions by \citet{1995PASJ...47..251S} demonstrated these small transient brightenings had an energy range of $10^{27} - 10^{29}$ ergs and a similar frequency distribution with larger hard X-rays flares.
Since the microflaring events in active regions couldn't provide enough energy to heat the solar corona, new studies focused on the small events in the quiet Sun.
The microflares in the quiet corona were investigated by \citet{1997ApJ...488..499K}, who found these events have an occurrence rate of one per three seconds extrapolated over the solar surface and the thermal energy of every event is $10^{25} - 10^{26}$ ergs. 
\citet{2000ApJ...535.1047A} detected 281 flare-like events in the energy range of $10^{24} - 10^{26}$ ergs in the 171 \AA\ and 
195 \AA\ channels with the \textit{Transition Region And Coronal Explorer} (\emph{TRACE}; \cite{1999SoPh..187..229H}) satellite. They found that these small flare-like events were miniature versions of larger flares in soft and hard X-rays and that their energy flux was  a factor $\approx 300$ below the quiet coronal heating requirements. With the \textit{Solar and Heliospheric Observatory} (SOHO; \cite{1995SoPh..162....1D}) observations, \citet{2002ApJ...568..413B} estimated that the energy input of EUV microflares was about  
$10\%$ of the radiative output in the same region. \citet{2008ApJ...677..704H} analyzed the thermal and non-thermal properties of 
microflares which had a mean thermal energy of $10^{28}$ ergs at the time of peak emission in $6 - 12$ keV. 
The characteristics of microflares in coronal holes and in quiet regions are studied by \citet{2011A&A...529A..21K} and these microflares were found to share common characteristics to active region flares. With combined
the \textit{Solar Dynamics Observatory} (\emph{SDO}; \cite{2012SoPh..275....3P}) and \textit{Reuven Ramaty High Energy Solar Spectroscopic Imager} (\emph{RHESSI}; \cite{2002SoPh..210....3L}) observations, \citet{2014ApJ...789..116I} presented that 8 of 10 microflares studied were fitted by a uniform differential 
emission measure profile.

Since \citet{1988ApJ...330..474P} suggested a nanoflare coronal heating theory, the efforts to search the smallest flare events has 
never been stopped. However, due to the small energy and the resolving 
capability of scientific instruments, the search and study of small flare-like events have never been satisfactory. It is necessary to
examine the characteristics of microflares more carefully with higher resolution data.
The launch of \emph{SDO} provides us a brand new chance to study microflares with more elaborate temporal and spatial resolution. In this paper, we name these small-scale impulsive transient events observed in 171 \AA\ as
EUV microflares, and focus on their statistical properties, spatial distribution as well as the relationship
with small-scale magnetic fields.

\section{Observations and data analysis}

The Atmospheric Imaging Assembly (AIA; \cite{2012SoPh..275...17L}) on board the \emph{SDO}
provides high resolution solar images with a pixel size of 0$\arcsec$.6 and a cadence 
of 12 s, while the Helioseismic and Magnetic Imager (HMI; \cite{2012SoPh..275..207S, 2012SoPh..275..229S}) provides
0$\arcsec$.5 resolution and 45 s cadence line-of-sight magnetograms uninterruptedly. In this study, we adopt
AIA 171 \AA\ EUV data and HMI line-of-sight magnetograms
on 2010 July 20, when there were no active regions on the solar disk. All the data we used have been applied differential 
rotation correction.

Firstly we select a quiet region with a field of view (FOV) of  $960{\arcsec} \times 1068{\arcsec}$ around solar center, and then we visually track the microflares one by one in the region. Finally 
we investigate the position, lifetime, area, brightness of the microflares respectively and determine the occurrence rate of them in latitudinal direction. Totally 10794 microflares are traced in this work.
A microflare is identified as follows: (1) We first employ the EUV movies to find microflare-like events with impulsive enhancements as 
candidates. (2) 
We select a square region for each event as carefully as possible to make sure that it can cover the target and doesn't include any pixels belong to another one. Most of the time, a square region works well, considering that the occurring rate of microflares is only $7.2 \times 10^{-4}$ Mm$^{-2}$ hr$^{-1}$ (10794 microflares in the FOV within 24 hr). However, there are still a few special cases when two events come too close. Under  these circumstances, a square region will include pixels from another event, we refer to an irregular shape, which will be shown in Figure~\ref{fig6}.
(3) We define an intensity enhancement as the difference of the integrated value in the square at the peak time and that at the time just before the event begins.  A microflare event is confirmed if the intensity enhancement exceeds 3$\sigma$, where $\sigma$ is the standard deviation of the integrated values in the square of each snapshot in the event process. 
Once a microflare is confirmed, its lifetime, position, area, and brightness are recorded. The position is determined by the brightest point of a microflare at the peak time. In previous studies, the area was often determined by the pixels exceeding a certain brightness level, which would fail for weak brightenings. Considering the sharp gradients between the microflare and its background, we obtain the area of a microflare by summing up the pixels whose value surpass the mean value in the square at the peak time. To remove the projection effect, 
the projected area is corrected to the real area by a $\cos(\alpha)$ factor, where $\alpha$ corresponds to 
the heliocentric angle. The brightness is averaged over the area and normalized to that of the quiet Sun ($I_{0}$), which is defined as the mean value of the whole image.

\section{Results}

We trace 10794 microflares totally in the chosen quiet region. 
For these microflares, their position, lifetime, area, brightness and the occurrence rate 
in latitudinal direction are investigated.
The relationship between the microflares 
and the underlying magnetic fields are also studied. 

\begin{figure}
\begin{center}
\includegraphics
[viewport=11 11 480 310, width=\textwidth]{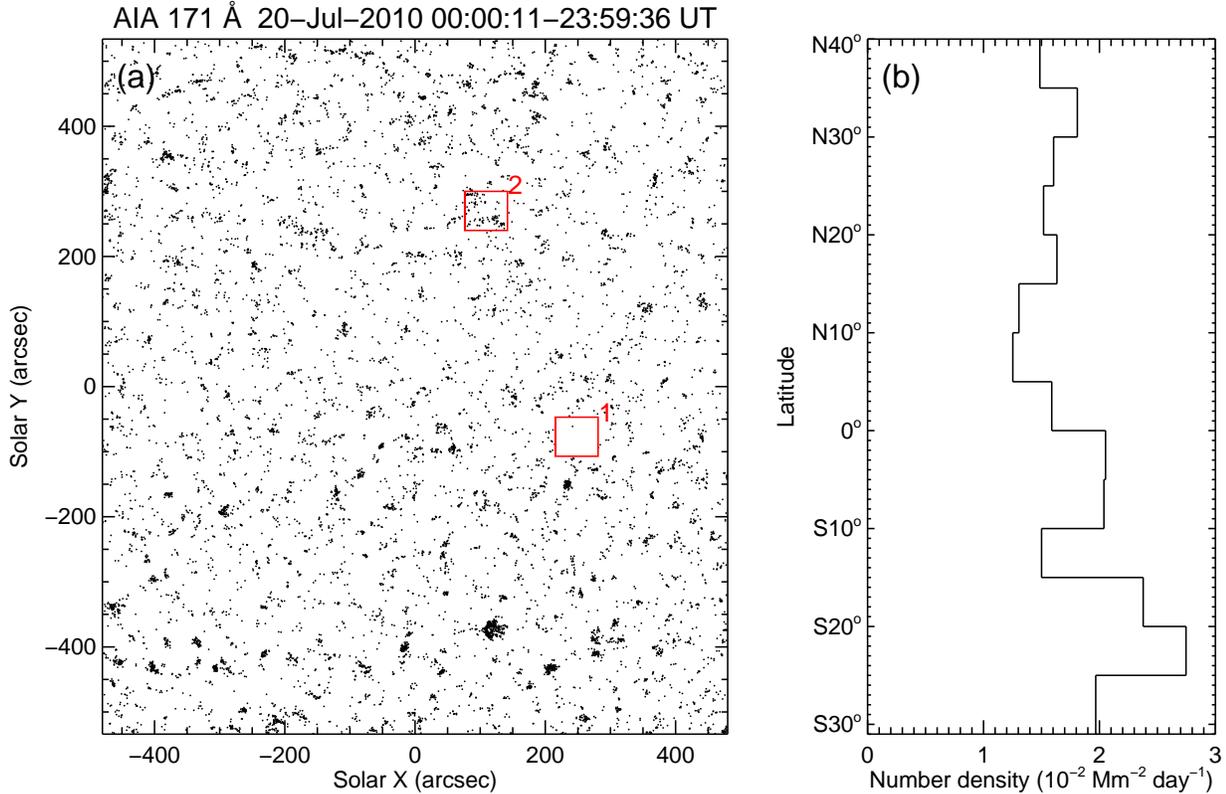} 
\end{center}
\caption{
{Distribution of the microflares obtained by the \emph{SDO}}$/$AIA.
\emph{Panel (a)}: distribution of microflares on the solar disk.
Each point indicates
a single microflare. Windows ``1" and ``2" show microflare-scarce and microflare-concentrated regions, respectively.
\emph{Panel (b)}: distribution of microflares along 
the latitudinal direction.}
\label{fig1}
\end{figure}

\subsection{Statistical results of microflares}

The left panel of Figure~\ref{fig1} shows the distribution of all the 10794 microflares over the solar disk. 
Each point indicates a 
single microflare, neglecting its size, brightness, etc. There exist microflare-concentrated areas, as well as ``void" areas without microflares. As a result, the microflares form visible network 
patterns on the solar disk. Although they occur unevenly in local regions, on the whole, the microflares still have an uniform distribution on the solar disk. We pick two typical regions with the FOV of $60{\arcsec} \times 60{\arcsec}$ (outlined by red windows ``1" and ``2" in panel (a)).  Within 24 hr, only one microflare occurred in window ``1" and 78 microflares appeared in window ``2". The occurrence rate of microflares are $2.12 \times 10^{-5}$ Mm$^{2}$ hr$^{-1}$ and $1.64 \times 10^{-3}$ Mm$^{2}$ hr$^{-1}$, respectively.
The microflares occupy about $0.05\%$ of the solar surface and their occurrence rate is $1.72 \times 10^{-2}$~Mm$^{-2}$~day$^{-1}$. The right panel 
of Figure~\ref{fig1} shows the distribution of microflares along the latitudinal direction. 
Because the center of the FOV lies in the north of the solar equator on July 20, the area in the south hemisphere is less than that in the northern hemisphere.
With 5558 microflares in the southern hemisphere and 5236 in the northern one, we get an occurrence rate of
$2.05 \times 10^{-2}$ Mm$^{2}$ day$^{-1}$ in the southern hemisphere, and $1.48 \times 10^{-2}$ Mm$^{2}$ day$^{-1}$ in the northern hemisphere. In other words, over the whole Sun, the number of microflares occurred in the southern hemisphere is 720 more than those occurred in the northern hemisphere for every hour.

\begin{figure}
\begin{center}
\includegraphics
[viewport=40 50 285 280, width=\textwidth]{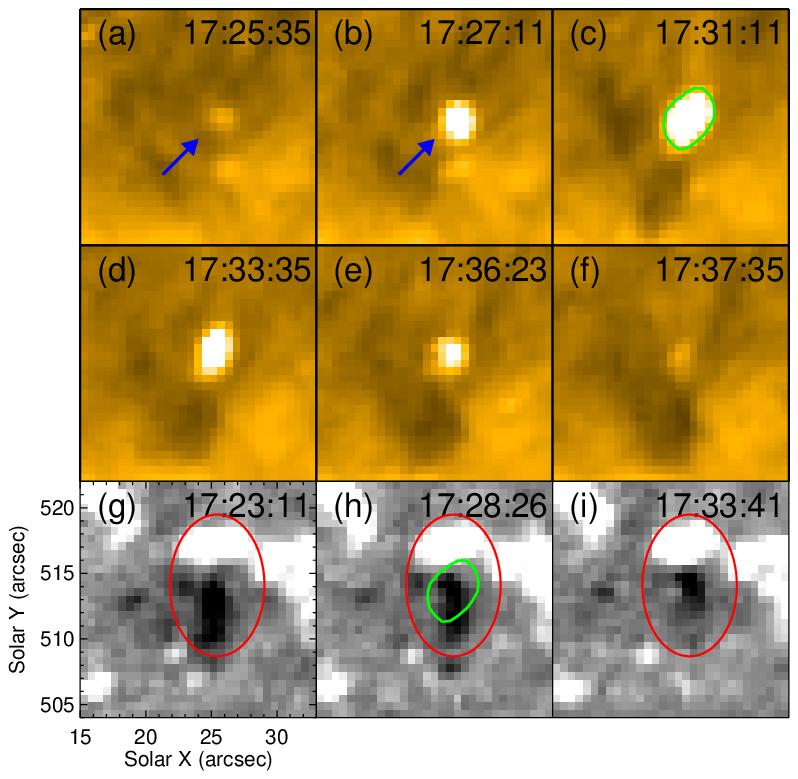} 
\end{center}
\caption{
\emph{Panels (a)--(f)}: AIA 171 \AA\ images showing the evolution of a microflare. The 
arrows in panel (a) and panel (b) indicate the location of the microflare.
\emph{Panels (g)--(i)}: corresponding magnetograms showing the cancellation of opposite polarities, as shown in the red ellipses.
The contour of the microflare in panel (c) is overlaid in panel (h) with the green curve.} \label{fig2}
\end{figure}

Figure~\ref{fig2} provides the evolution process of a microflare from appearance to disappearance.  The arrows
in panels (a) and (b) indicate where the microflare is located. The microflare started at  17:25 UT,  
then grew up and reached its peak at 17:31 UT . After that, it decayed gradually, and disappeared finally
at 17:37 UT. The brightness, area and lifetime of this microflare are 7.4 $I_{0}$, 5.2 Mm$^{2}$ and 12.0 min, respectively.
As can be seen from the three bottom 
panels, the opposite polarity magnetic fields canceled 
with each other (outlined by red ellipses). We overlay the microflare on the magnetogram by green contour curve in panel (h)
and find that the microflare lies mainly on the cancellation site of opposite polarities. From 17:25 UT to 17:37 UT, the total unsigned magnetic flux decreased by $2.0 \times 10^{18}$ Mx due to the cancellation.

\begin{figure}
\begin{center}
\includegraphics
[viewport=0 20 490 540,width=\textwidth]{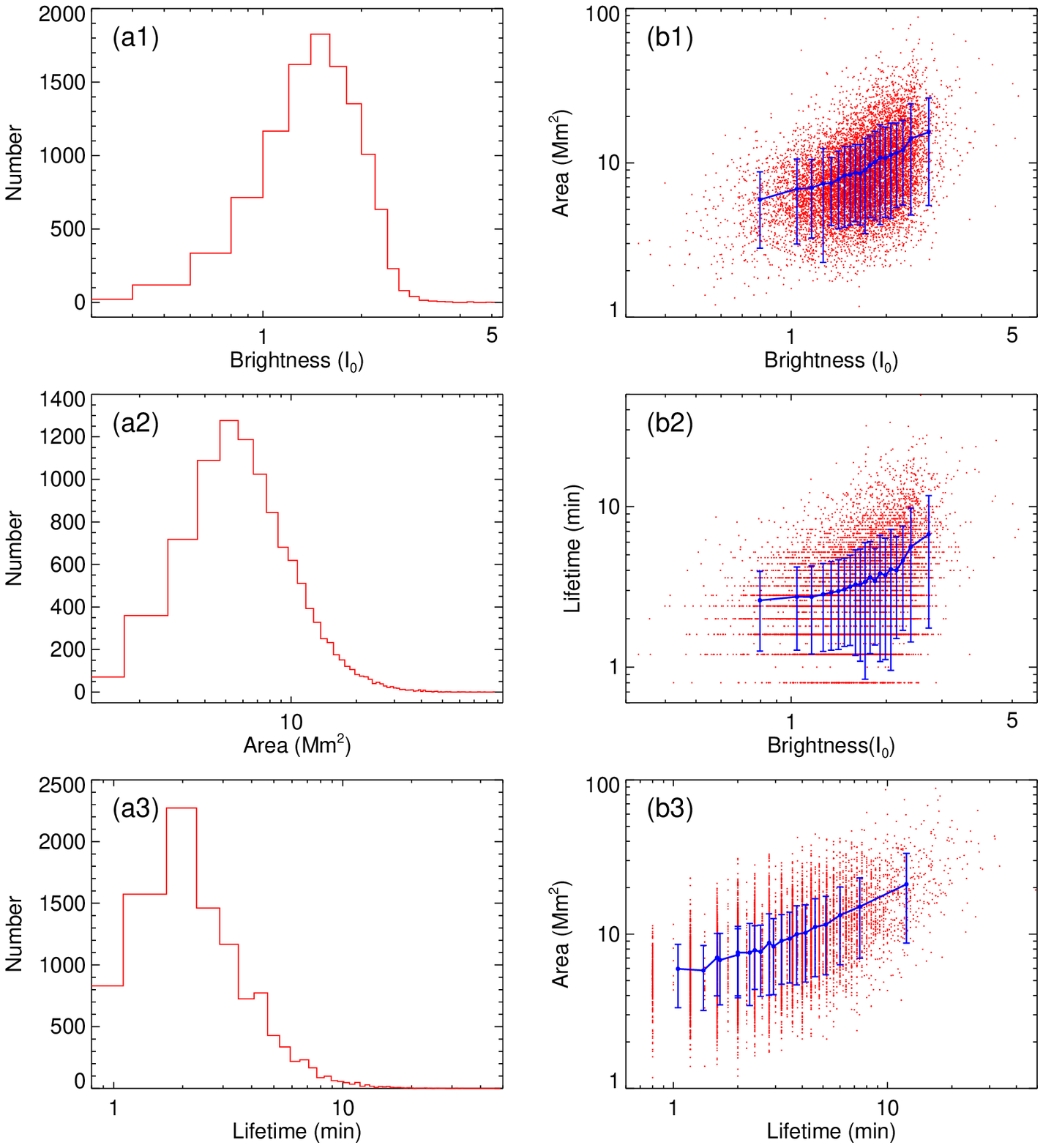}
\end{center}
\caption{
\emph{Panel (a1)}: histogram of brightness with a binsize of 0.2 $I_{0}$. 
\emph{Panel (a2)}: histogram of area with a binsize of 1.0 Mm$^{2}$.
\emph{Panel (a3)}: histogram of lifetime with a binsize of 36 s.
\emph{Panels (b1), (b2) and (b3)}: scatter plots of brightness, lifetime, and
area with each other (red symbols) and of sort-grouped points
with error bars (blue symbols).} \label{fig3}
\end{figure}

The statistical results of brightness, area, and lifetime, together with their mutual correlation with each other, 
are shown in Figure~\ref{fig3}.
The brightness varies from 0.3~$I_{0}$ to 5.2~$I_{0}$. 
There is a peak located at 1.5 $I_{0}$. As the brightness $I_{0}$ is obtained by normalizing the full disk, a microflare which takes place
in dark background may have a brightness 
smaller than 1 $I_{0}$.
As revealed by panel (a2), 
the area of microflares covers from 1.2 Mm$^{2}$ to 88.2 Mm$^{2}$, and is most concentrated in the level of 
5.0 Mm$^{2}$. 
Similar to panels (a1) and (a2), the histogram of lifetime is plotted in panel (a3).  The microflares' lifetime spans from
48 s to 49.2 min, with a peak at 2 min. The mean values of the three parameters are 1.7~$I_{0}$, 9.6 Mm$^{2}$, 
and 3.6 min, respectively.

The scatter plots of brightness, area, and lifetime with one another are shown with red symbols 
in panels (b1), (b2) and (b3) of Figure~\ref{fig3}. The mutual correlations seem quite diffuse. To make the correlations more clear, we apply a 
``sort-group" method suggested by \citet{2009RAA.....9..933Z}. Take panel (b1) for example. Firstly, we sort
the area according to the brightness of individuals.  Secondly, the sorted data are divided into
19 groups with 540 elements each and one group with 534 elements.  Then we get 20 data points by assigning with 
the mean value of each group. Thirdly, the area and brightness of microflares are correlated with each other and plotted with blue symbols. The error bars are plotted with one standard deviation from the mean value of area in each group in panel (b1).
Although the three parameters have large deviations from the mean value in each group,
the mutual positive correlation of brightness, area, and lifetime with each other become clear after the trick. The statistical correlations reveal a tendency that the larger microflares have larger brightness and longer lifetimes.

\subsection{Relationship between microflares and small-scale magnetic fields}

\begin{figure}
\begin{center}
\includegraphics
[viewport=0 20 540 535,width=\textwidth]{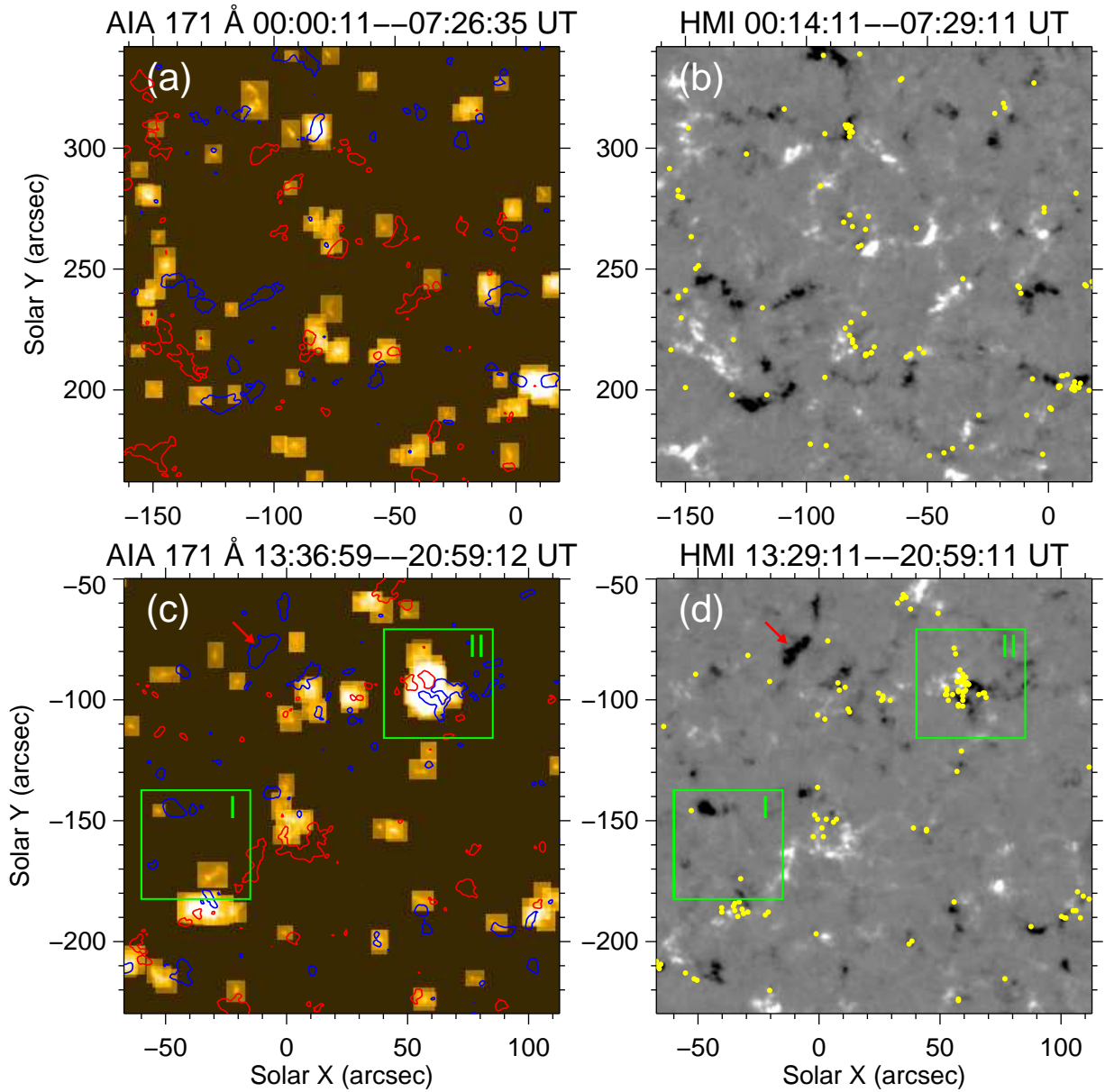}
\end{center}
\caption{
Two random picked regions of microflares and their corresponding 
magnetic fields.  Windows ``I" and ``II" outline two typical regions. One is scarce of microflares, while the other is prosperous.
The red and blue contour lines in the left panels 
represent $\pm$15 G of the magnetic fields, while 
each yellow point in the right panels stands for a single microflare. The red arrow in panel (d) indicates the uni-polarity network
magnetic fields.}
\label{fig4}
\end{figure}

To study the relationship between the microflares and the small-scale magnetic fields, we randomly select two regions, 
as shown in Figure~\ref{fig4}. The left two panels show all the microflares occurred in the two random picked regions within about 7.5 hr. For each region, every microflare during that period is recorded by a small square window. All the windows in the region are superimposed together to form a new image, as displayed in Figure 4(a) and 4(c).
The right two panels display the corresponding magnetic fields, which are obtained by averaging over 30 magnetograms with
a cadence of 15 min.
We overlay the magnetic contour with $\pm$15 G on the microflare images, and mark the microflares' positions
with yellow symbols on the magnetograms. It seems that most of the microflares occur above the network magnetic fields.
However, the microflares hardly take place in the uni-polarity network magnetic fields, as indicated by red arrows in panels (c) and (d) of Figure~\ref{fig4}.
Within the intranetwork area where the magnetic fields are weak, few microflares are seen. 
These microflares form 
some network-like patterns, coinciding with the corresponding magnetic network. 

\begin{figure}
\begin{center}
\includegraphics
[viewport=15 50 530 310,width=\textwidth]{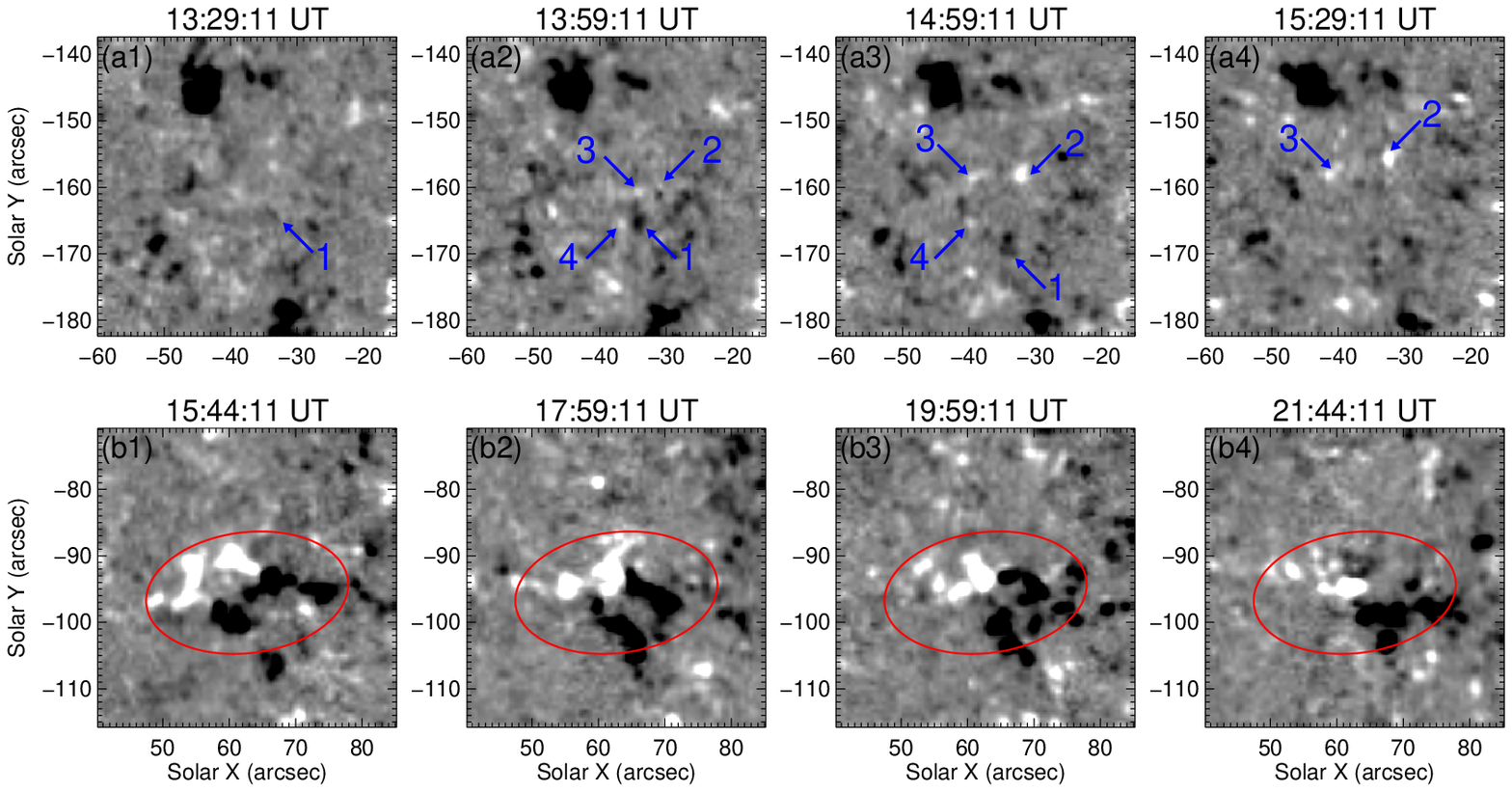}
\end{center}
\caption{
\emph{Panels (a1)--(a4)}: sequence of HMI magnetograms outlined by 
square ``I" in Figure~\ref{fig4} (d).
Arrows ``1", ``2", ``3" and ``4" denote four emerging magnetic elements.
\emph{Panels (b1)--(b4)}: similar to \emph{panels (a1)--(a4)}, but are 
relevant to the square ``II" in Figure~\ref{fig4} (d). The ellipses outline 
where the magnetic cancellations take place.}
\label{fig5}
\end{figure}

We carry out a detail check on the magnetic evolution in two small subregions marked by windows 
``I" and ``II" in panels (c)
and (d) of Figure~\ref{fig4}. Window ``I" is scarce of microflares, with an occurrence rate of 
$2.4 \times 10^{-4}$ Mm$^{-2}$ hr$^{-1}$, while microflares occur frequently in window ``II", with an occurrence rate of $4.8 \times 10^{-3}$ Mm$^{-2}$ hr$^{-1}$. The evolution sequences of magnetic fields are shown in Figure~\ref{fig5}, with window ``I"
in the top panels and  ``II" in the bottom panels. To decrease the noise level, three magnetograms with a 
cadence of 45 s are averaged
to form a new one. At 13:29 UT,  a magnetic element emerged, which is labeled
by arrow ``1" in panel (a1). Half an hour later, it had grown larger. Meanwhile, three more turned up, 
as shown with arrows ``2", ``3", and ``4"  in panel (a2). At 14:59 UT, these four elements grew larger and separated with each
other, which can be seen in panel (a3). In panel (a4), elements ``1" and ``4" could not be identified anymore. However, 
elements ``2" and ``3" still moved further and further.  In window ``II" where the microflares appear frequently, 
there are two pair of magnetic elements canceling with each other, as indicated with ellipses in panels (b1) -- (b4).

\begin{figure}
\begin{center}
\includegraphics
[viewport=40 80 500 430,width=\textwidth]{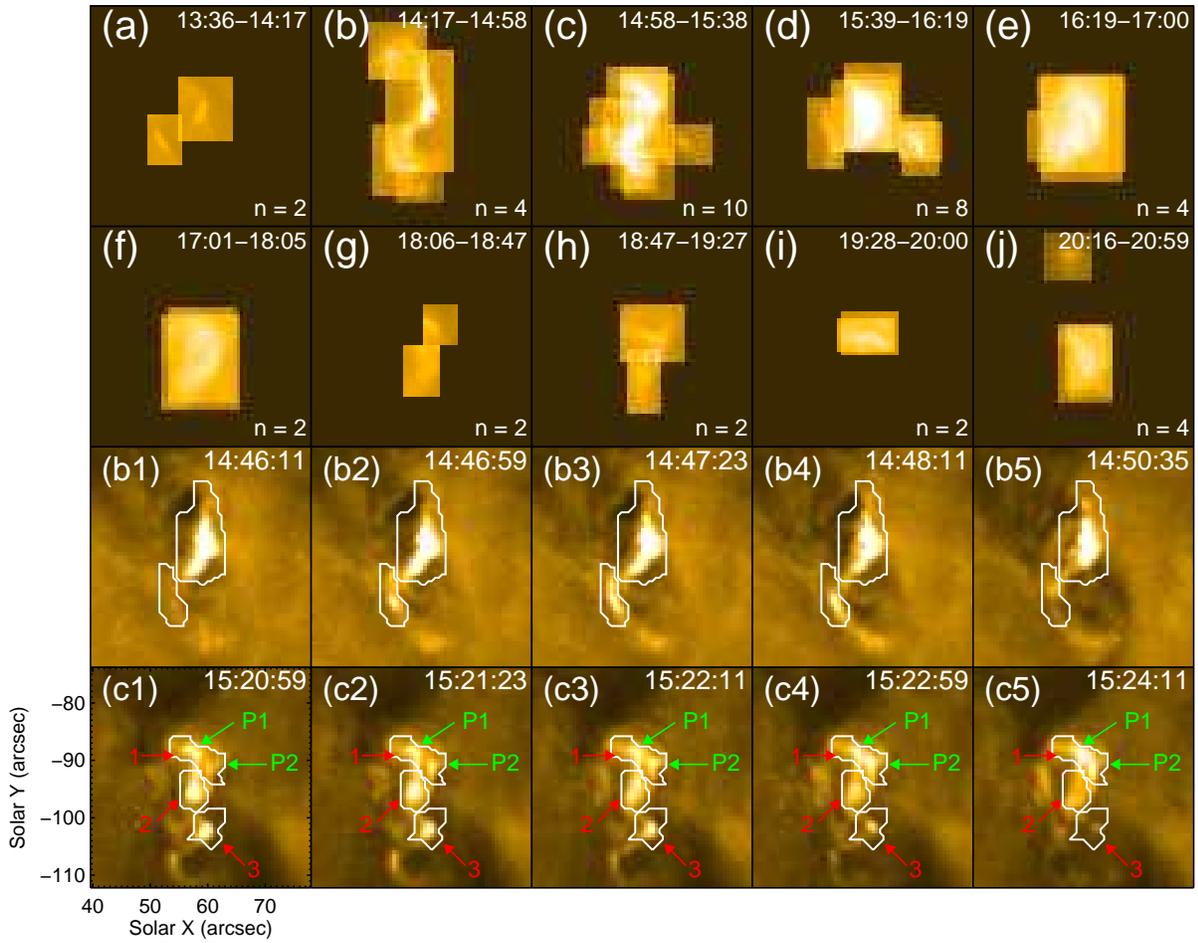}
\end{center}
\caption{
\emph{Panels (a)--(j)}: microflares occurred in region ``II" of Figure~\ref{fig4} (c), but within about 40 min for each panel.
The number of microflares that occurred for each time interval is shown in the  lower right corner of each panel.
\emph{Panels (b1)--(b5)}: sequence of EUV images which demonstrate how to pick up two microflares coming 
too close in panel (b).
\emph{Panels (c1)--(c5)}: Arrows ``1", ``2", and ``3" denote three microflares recorded in panel (c). Microflare ``1" is treated as
a single one since ``P1" and ``P2" merge and can't be distinguished with each other in their life cycles. The three microflares are tracked similar to those in panels (b1)--(b5).
}
\label{fig6}
\end{figure}

\section{Conclusion and discussion}

By using the high resolution 171 \AA\ images from AIA and line-of-sight magnetograms obtained by HMI on board the \emph{SDO}, we trace 10794 microflares in a quiet region with a FOV of $960\arcsec \times 1068\arcsec$. Their statistical properties, spatial distribution and relationship with
magnetic fields are investigated. The microflares distribute unevenly in local regions, but on the whole, they have an uniform distribution over the whole region. The occurrence rate is $1.72 \times 10^{-2}$ Mm$^{-2} $day$^{-1} $, i.e., $4.4\times10^{3}$ hr$^{-1}$ extrapolated over the whole Sun.
We explore microflares down to 48 s over time scale and to 1.2 Mm$^{2}$ in size. 
The average size of microflares is 9.6 Mm$^{2}$ and the average lifetime is 3.6 min.
By applying a ``sort-group" method, 
we find a mutual positive correlation of the microflares' brightness, area, and lifetime.  
The distribution of microflares exhibits 
network patterns, which are similar to and matched with corresponding magnetic network structures. After a detail check with the  magnetograms, we find that the microflares are more concentrated in the magnetic cancellation regions, and hardly take place in the intranetwork regions, where many new small-scale magnetic elements emerge.

Due to the superimposition of microflares in more than 7 hr in the left two panels of Figure~\ref{fig4}, it is difficult to  resolve individual microflares, especially in the dense regions like the region ``II".  So we dissemble the region ``II" in panel (c) of Figure~\ref{fig4} into 10 panels (Figures 6(a)-(j)) with shorter time intervals. The number of microflares occurred for each time interval is given in the bottom-right of each panel. As mentioned in Section 2, the method of picking up microflares with square regions will fail for very special cases. Panels (b1)--(b5) and (c1)--(c5) in Figure~\ref{fig6} show two special examples how we deal with these situations. In panel (b), there were two microflares coming close in space and in time. 
So we employ an irregular shape instead of a square one to pick up one microflare so as not to include pixels from the other, which is shown in panels (b1)--(b5). The methods of microflare identification and parameter determination are the same to those described in Section 2. As shown in panels (c1)--(c5), two impulsive events ``P1" and ``P2"  arose and merged with each other in their lifetime. As they couldn't be distinguished and tracked individually, ``P1" and ``P2" were treated as one microflare
and investigated as a whole. Together with microflares ``2" and ``3", they are tracked with irregular shapes as mentioned above.

Thanks to the high-resolution data provides by \emph{SDO}, we can distinguish microflares with lifetime as short as 48 s and with area as small as 1.2 Mm$^{2}$. 
As a result, we observe 3.7 times microflares observed by \cite{1997ApJ...488..499K} (1200 hr$^{-1}$). 
\citet{2001SoPh..198..347Z} presented a statistical study of coronal BPs and found the average size of a BP is 110 Mm$^{2}$.
\citet{2011A&A...529A..21K} identified 7 microflares in quiet regions whose area ranges from 8.4 Mm$^{2}$ to 22 Mm$^{2}$. 
In our study, the area of microflares covers from 1.2 Mm$^{2}$ to 88.2 Mm$^{2}$ with a peak at 5.0 Mm$^{2}$, which is much smaller than that in \citet{2001SoPh..198..347Z}, but similar to that in \citet{2011A&A...529A..21K}. The typical lifetime of microflares was found to be 5 -- 20 min \citep{1981SoPh...69...77H, 1990ApJ...352..333H, 1997ApJ...488..499K, 2002ApJ...568..413B, 2011A&A...529A..21K}. However, more than $80\%$ of the microflares we observed have a lifetime less than 5 min.
The statistical correlations reveal a tendency that the larger microflares have higher brightness and longer lifetimes.
However, the mutual positive correlation between brightness, area and lifetime may be not a welcome sign for smallest events detection.

Our results reveal that the microflares are prosperous in the magnetic cancellation regions of network boundaries and 
wilted in the intranetwork regions. 
Figure~\ref{fig5} suggests such a scenario: new magnetic elements emerge within the intranetwork region, move apart, and then interact with pre-existing network magnetic fields. The reconnection between newly emerged magnetic fields and pre-existing ones results in the microflares. This scenario is consistent with the converging flux model suggested by \citet{1994ApJ...427..459P}, which described how two approaching opposite polarities caused a null point  that arose into the corona and formed a BP by magnetic reconnection.
The ubiquity of microflares across the solar surface is crucial to the coronal heating. The network patterns exhibited by microflares' distribution are similar to the magnetic network structures. However, whether the ``void" area without microflares is a true microflare-free zone or just a result of detection limitation can not be determined yet. Higher resolution observations are needed to make a more careful examination.

Let's make a rough estimation of the energy flux of EUV microflares in our observations. The occurrence rate is about  $4.4\times10^{3}$ hr$^{-1}$
extrapolated over the whole Sun. Taking the upper limit energy of microflares, i.e., 10$^{27}$ erg, then we get an average energy flux of about
$2.0 \times 10^{4}$ erg s$^{-1} $cm$^{-2}$. However, the energy flux required for the coronal heating in the quiet Sun is about 
$3 \times 10^{5}$ erg s$^{-1} $cm$^{-2}$ \citep{1977ARA&A..15..363W}. Our result is still a factor of $\sim 15$ below the coronal heating requirement. We suggest there may exist a sea of nanoflares \citep{2013ApJ...770L...1T} beneath the detection limit. There are also some studies \citep{2000ApJ...535.1047A, 2000ApJ...529..554P} suggesting that the events with picoflare energies may contribute to the coronal heating because of the insufficient energy of nanoflares. Besides, the Alfv{\'e}n waves \citep{1947MNRAS.107..211A}, which can transport magneto-convective
energy upwards, have been invoked as another promising mechanism to heat the solar corona \citep{2011ApJ...736....3V}. \citet{2007Sci...318.1574D} have observed  Alfv{\'e}n waves with sufficient energy in the chromosphere possibly to heat the solar corona. Ubiquitous Alfv{\'e}n motions have also been observed in the transition region and in the corona by \citet{2011Natur.475..477M} with the \emph{SDO} satellite, although the generation and dissipation of these waves are still not clear. The ubiquitous rotating network magnetic fields (RNFs) and EUV cyclones in the quiet Sun suggested by \citet{2011ApJ...741L...7Z} may be another effective coronal heating mechanism. The field lines are braided by continuous developments of RNFs and then magnetic reconnection occurs in the braiding fields. Thus the stored energy can be released to heat the corona, while the EUV waves following the cyclones transport energy to other places.

\bigskip

We thank the \emph{SDO}/AIA and HMI teams for providing data.
This work is funded by the National Natural Science Foundations
of China (11221063, 11203037, 11303050), the
CAS Project KJCX2-EW-T07, the National Basic Research Program of
China under grant 2011CB811403, and the Strategic Priority Research Program -- The Emergence of Cosmological Structures of 
the Chinese Academy of Sciences (No. XDB09000000).


\begin{thebibliography}{}

\bibitem[Alfv{\'e}n(1947)]{1947MNRAS.107..211A} Alfv{\'e}n, H.\ 1947, 
\mnras, 107, 211

\bibitem[Aschwanden(2004)]{2004psci.book.....A} Aschwanden, M.~J.\ 2004, 
Physics of the Solar Corona. An Introduction (Chichester: Praxis)

\bibitem[Aschwanden et al.(2000)]{2000ApJ...535.1047A} Aschwanden, M.~J., 
Tarbell, T.~D., Nightingale, R.~W., et al.\ 2000, \apj, 535, 1047 

\bibitem[Benz 
\& Krucker(2002)]{2002ApJ...568..413B} Benz, A.~O., \& Krucker, S.\ 2002, \apj, 568, 413

\bibitem[Berger et al.(2007)]{2007ApJ...661.1272B} Berger, T.~E., Rouppe 
van der Voort, L., L\"ofdahl, M.\ 2007, \apj, 661, 1272

\bibitem[Berghmans et 
al.(2001)]{2001A&A...369..291B} Berghmans, D., McKenzie, D., \& Clette, F.\ 2001, \aap, 369, 291

\bibitem[Chandrashekhar et al.(2013)]{2013SoPh..286..125C} Chandrashekhar, 
K., Krishna Prasad, S., Banerjee, D., Ravindra, B., 
\& Seaton, D.~B.\ 2013, \solphys, 286, 125 


\bibitem[De Pontieu et al.(2007)]{2007Sci...318.1574D} De Pontieu, B., 
McIntosh, S.~W., Carlsson, M., et al.\ 2007, Science, 318, 1574

\bibitem[Domingo et al.(1995)]{1995SoPh..162....1D} Domingo, V., Fleck, B., 
\& Poland, A.~I.\ 1995, \solphys, 162, 1 

\bibitem[Doschek et al.(2010)]{2010ApJ...710.1806D} Doschek, G.~A., Landi, 
E., Warren, H.~P., \& Harra, L.~K.\ 2010, \apj, 710, 1806

\bibitem[Golub et al.(1974)]{1974ApJ...189L..93G} Golub, L., Krieger, 
A.~S., Silk, J.~K., Timothy, A.~F., \& Vaiana, G.~S.\ 1974, \apjl, 189, L93 

\bibitem[Golub 
\& Pasachoff(1997)]{1997soco.book.....G} Golub, L., \& Pasachoff, J.~M.\ 1997, The Solar Corona (Cambridge: Cambridge Univ. Press)

\bibitem[Habbal 
\& Withbroe(1981)]{1981SoPh...69...77H} Habbal, S.~R., \& Withbroe, G.~L.\ 1981, \solphys, 69, 77

\bibitem[Habbal et al.(1990)]{1990ApJ...352..333H} Habbal, S.~R., Withbroe, 
G.~L., \& Dowdy, J.~F., Jr.\ 1990, \apj, 352, 333 

\bibitem[Handy et al.(1999)]{1999SoPh..187..229H} Handy, B.~N., Acton, 
L.~W., Kankelborg, C.~C., et al.\ 1999, \solphys, 187, 229 

\bibitem[Hannah et al.(2008)]{2008ApJ...677..704H} Hannah, I.~G., Christe, 
S., Krucker, S., et al.\ 2008, \apj, 677, 704 

\bibitem[Harrison(1997)]{1997SoPh..175..467H} Harrison, R.~A.\ 1997, 
\solphys, 175, 467

\bibitem[Hudson(1991)]{1991SoPh..133..357H} Hudson, H.~S.\ 1991, \solphys, 
133, 357

\bibitem[Inglis 
\& Christe(2014)]{2014ApJ...789..116I} Inglis, A.~R., \& Christe, S.\ 2014, \apj, 789, 116

\bibitem[Kamio et al.(2011)]{2011A&A...529A..21K} Kamio, S., Curdt, W., Teriaca, L., \& Innes, D.~E.\ 2011, \aap, 529, A21 

\bibitem[Krucker et al.(1997)]{1997ApJ...488..499K} Krucker, S., Benz, 
A.~O., Bastian, T.~S., \& Acton, L.~W.\ 1997, \apj, 488, 499 

\bibitem[Lemen et al.(2012)]{2012SoPh..275...17L} Lemen, J.~R., Title, 
A.~M., Akin, D.~J., et al.\ 2012, \solphys, 275, 17

\bibitem[Levine(1974)]{1974ApJ...190..457L}
Levine, R.~H.\ 1974, \apj, 190, 457

\bibitem[Lin et al.(1984)]{1984ApJ...283..421L} Lin, R.~P., Schwartz, 
R.~A., Kane, S.~R., Pelling, R.~M., \& Hurley, K.~C.\ 1984, \apj, 283, 421

\bibitem[Lin et al.(2002)]{2002SoPh..210....3L} Lin, R.~P., Dennis, B.~R., 
Hurford, G.~J., et al.\ 2002, \solphys, 210, 3 

\bibitem[McIntosh et al.(2011)]{2011Natur.475..477M} McIntosh, S.~W., de 
Pontieu, B., Carlsson, M., et al.\ 2011, \nat, 475, 477 

\bibitem[Parker(1988)]{1988ApJ...330..474P} Parker, E.~N.\ 1988, \apj, 330, 
474 

\bibitem[Parnell 
\& Jupp(2000)]{2000ApJ...529..554P} Parnell, C.~E., \& Jupp, P.~E.\ 2000, \apj, 529, 554 


\bibitem[Parnell(2002)]{2002ESASP.505..231P} Parnell, C.~E.\ 2002, SOLMAG 
2002.~Proceedings of the Magnetic Coupling of the Solar Atmosphere 
Euroconference, 505, 231

\bibitem[Parnell 
\& De Moortel(2012)]{2012RSPTA.370.3217P} Parnell, C.~E., \& De Moortel, I.\ 2012, Royal Society of London Philosophical Transactions Series A, 370, 3217

\bibitem[Pesnell et al.(2012)]{2012SoPh..275....3P} Pesnell, W.~D., 
Thompson, B.~J., \& Chamberlin, P.~C.\ 2012, \solphys, 275, 3

\bibitem[Porter et al.(1987)]{1987ApJ...323..380P} Porter, J.~G., Moore, 
R.~L., Reichmann, E.~J., Engvold, O., 
\& Harvey, K.~L.\ 1987, \apj, 323, 380

\bibitem[Priest et al.(1994)]{1994ApJ...427..459P} Priest, E.~R., Parnell, 
C.~E., \& Martin, S.~F.\ 1994, \apj, 427, 459

\bibitem[Riethm{\"u}ller et al.(2010)]{2010ApJ...723L.169R} 
Riethm{\"u}ller, T.~L., Solanki, S.~K., Mart{\'{\i}}nez Pillet, V., et al.\ 
2010, \apjl, 723, L169 

\bibitem[Romano et al.(2012)]{2012SoPh..280..407R} Romano, P., Berrilli, 
F., Criscuoli, S., et al.\ 2012, \solphys, 280, 407 

\bibitem[Scherrer et al.(2012)]{2012SoPh..275..207S} Scherrer, P.~H., 
Schou, J., Bush, R.~I., et al.\ 2012, \solphys, 275, 207

\bibitem[Schou et al.(2012)]{2012SoPh..275..229S} Schou, J., Scherrer, 
P.~H., Bush, R.~I., et al.\ 2012, \solphys, 275, 229

\bibitem[Shimizu(1995)]{1995PASJ...47..251S} Shimizu, T.\ 1995, \pasj, 47, 
251

\bibitem[Testa et al.(2013)]{2013ApJ...770L...1T} Testa, P., De Pontieu, 
B., Mart{\'{\i}}nez-Sykora, J., et al.\ 2013, \apjl, 770, L1 

\bibitem[Utz et 
al.(2009)]{2009A&A...498..289U} Utz, D., Hanslmeier, A., M{\"o}stl, C., et al.\ 2009, \aap, 498, 289

\bibitem[Vaiana et al.(1973)]{1973ApJ...185L..47V} Vaiana, G.~S., Davis, 
J.~M., Giacconi, R., et al.\ 1973, \apjl, 185, L47 

\bibitem[van Ballegooijen et al.(2011)]{2011ApJ...736....3V} van 
Ballegooijen, A.~A., Asgari-Targhi, M., Cranmer, S.~R., 
\& DeLuca, E.~E.\ 2011, \apj, 736, 3 

\bibitem[Withbroe 
\& Noyes(1977)]{1977ARA&A..15..363W} Withbroe, G.~L., \& Noyes, R.~W.\ 1977, \araa, 15, 363

\bibitem[Zakharov et 
al.(2005)]{2005A&A...437L..43Z} Zakharov, V., Gandorfer, A., Solanki, S.~K., L\"ofdahl, M.\ 2005, \aap, 437, L43

\bibitem[Zhang et al.(2001)]{2001SoPh..198..347Z} Zhang, J., Kundu, M.~R., 
\& White, S.~M.\ 2001, \solphys, 198, 347

\bibitem[Zhang 
\& Liu(2011)]{2011ApJ...741L...7Z} Zhang, J., \& Liu, Y.\ 2011, \apjl, 741, L7 

\bibitem[Zhao et al.(2009)]{2009RAA.....9..933Z} Zhao, M., Wang, J.-X., 
Jin, C.-L., 
\& Zhou, G.-P.\ 2009, Research in Astronomy and Astrophysics, 9, 933 


\end{thebibliography}
\end{document}